# PHASE AWARE EAR-CONDITIONED LEARNING FOR MULTI-CHANNEL BINAURAL SPEAKER SEPARATION

*Ruben Johnson Robert Jeremiah, Peyman Goli, Steven van de Par*

Acoustics Group, Department of Medical Physics and Acoustics, Cluster of Excellence Hearing4all, University of Oldenburg, Oldenburg, Germany

## ABSTRACT

Separating competing speech in reverberant environments requires models that preserve spatial cues while maintaining separation efficiency. We present a Phase-aware EAr-conditioned speaker Separation network using eight microphones (PEASE-8) that consumes complex STFTs and directly introduces a raw-STFT input to the early decoder layer, bypassing the entire encoder pathway to improve reconstruction. The model is trained end-to-end with an SI-SDR–based objective against direct-path ear targets, jointly performing separation and dereverberation for two speakers in a fixed azimuth, eliminating the need for permutation invariant training. On spatialized two-speaker mixtures spanning anechoic, reverberant, and noisy conditions, PEASE-8 delivers strong separation and intelligibility. In reverberant environments, it achieves 12.37 dB SI-SDR, 0.87 STOI, and 1.86 PESQ at T60 ≈ 0.6s, while remaining competitive under anechoic conditions.

## 1. INTRODUCTION

Separating overlapping speech in noisy and reverberant conditions remains a core challenge in audio signal processing. Deep neural networks (DNNs) have significantly improved monaural and binaural speaker separation, but these systems often struggle in adverse acoustic conditions due to limited directional cues [1]. Incorporating additional microphones increases spatial sampling. This enables richer exploitation of inter-channel level differences, as well as time-delay cues, which can substantially improve intelligibility [2].

Here we will investigate speaker separation in the context of hearing aids, which often carry multiple microphones at each ear. Although it is not the focus of this study, such speaker separation will allow retrospective processing of each speaker to have optimal spatial and monaural properties, fostering optimal speech intelligibility.

Recent works in multi-microphone speech separation has explored both time-domain (TD) and time-frequency (TF) domain approaches. Methods include Multi-Channel Conv-TasNet and its beam-forming variants (often referred to as Beam-TasNet) [3], which integrates learned spatial filtering with MVDR post-processing. Filter-and-sum networks, such as FaSNet [4] and iFaSNet [5], explicitly predict per-channel filters before summation, achieving strong separation quality in multi-microphone scenarios. Transformer and cross-attention models learn spatial representations end-to-end without hand-crafted features, but typically require permutation-invariant training (uPIT), which resolves source order ambiguities yet limits explicit control over output identity [6].

While many approaches discard phase information and rely solely on magnitude spectra, phase-aware models that retain complex short-time Fourier transform (STFT) features provide superior spatial disambiguation and reconstruction fidelity [7], [8]. Our ear-conditioned design propagates spatial cues to the decoder through skip connections, preserving spatial information for end-to-end separation

Building on our Phase-aware EAr-conditioned Separation Network (PEASE-Net) [9], we introduce PEASE-8, which employs a residual CNN encoder, multi-head attention, and per-source decoders with novel input-to-decoder skip connections bypassing the entire encoder. This work extends the framework to eight-channel microphone arrays. Unlike U-Net [10] or ResNet-style networks [11], PEASE-Net's skip connections project raw STFT features directly to ear-specific decoder heads, preserving spatial information for end-to-end separation. By incorporating three external microphones per hearing aid alongside in-ear channels, we scale this architecture to exploit higher-order spatial information while maintaining PIT-free operation with fixed azimuth targets. We note that using ear-conditioned targets implies the proposed method aims to extract dereverberated speakers from noisy and reverberated mixtures. Indeed, informal listening confirms this and has potential positive implications for speech intelligibility for hearing-impaired listeners.

## 2. METHODOLOGY

### 2.1. Problem Formulation

We employ an eight-channel microphone array comprising two in-ear channels plus three external microphones per hearing aid. The acoustic scene contains two speech sources and additive noise. The eight-channel observation at time $t$ is expressed as,

$$x(t) = \sum_{s=1}^{2} h_s(t) * s_s(t) + h_n(t) * n(t) \quad (1)$$

where $x(t)$ is the eight-channel observation vector, $s_s$ denotes the clean speech signal from the source $s$, $n(t)$ is the additive noise and $h_s(t)$ and $h_n(t)$ represent the corresponding eight-channel room impulse responses from each source and noise position to all microphones.

$$X(t,f) = [X_1(t,f), X_2(t,f), \cdots, X_8(t,f)]^T \in \mathbb{C}^8 \quad (2)$$

For each time frame t, we construct a temporal context window spanning 2τ+1 frames,

$$Y_t = [X_{t-\tau}, \cdots, X_{t-1}, X_t, X_{t+1}, \cdots, X_{t+\tau}] \quad (3)$$

The network estimates each speaker's direct-path complex STFT and reconstructs waveforms via iSTFT.

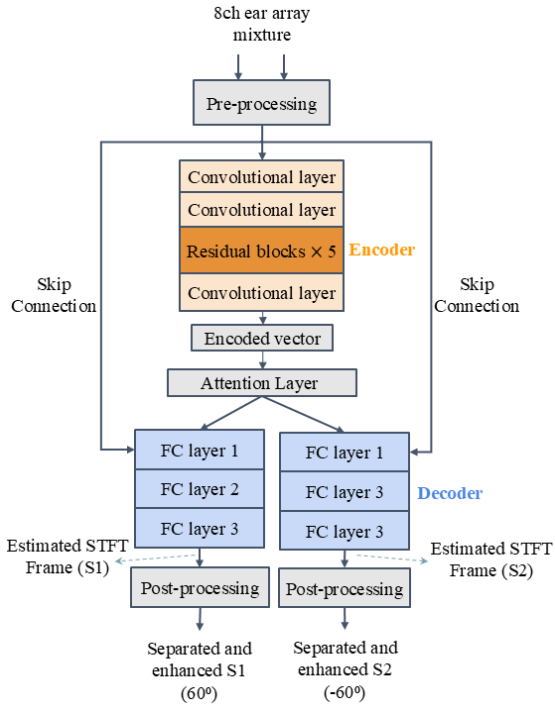

Fig. 1. Architecture of the proposed PEASE-8

## 2.2. Architecture

PEASE-8 extends PEASE-Net to eight microphones while preserving its phase-aware, ear-conditioned design as shown in Fig. 1. It ingests concatenated real/imaginary complex STFTs from all channels over a temporal context window, operating directly in the TF domain (no mask constraints) to exploit both spectral content and inter-channel spatial cues. The encoder performs multi-channel fusion via early spectral down-sampling with temporal coherence, followed by residual blocks; a temporal convolution extracts center-frame features for global processing. A multi-head attention layer models long-range spectral interactions for separation.

The distinguishing feature is an ear-conditioned skip connection that bypasses the encoder: raw eight-channel STFTs are split into left/right groups, linearly projected, and injected into the first decoder layer, creating spatially biased embeddings aligned with the target azimuths. Two parallel, independent decoders then produce per-source complex spectrograms. This deterministic ear conditioning fixes source–output assignment, eliminates PIT, and leverages the array's full spatial diversity.

## 2.3. Training Objective

PEASE-8 is trained end-to-end to directly regress the complex STFT of each target speaker without PIT. We employ the scale-invariant signal-to-distortion ratio (SI-SDR) as the objective, which measures the fidelity of the estimated waveform relative to the clean reference while remaining invariant to scaling factors. Given the estimated waveform $\hat{s}_s$ and reference $s$, SI-SDR is defined as

$$SI-SDR(\hat{s}, s) = 10 log_{10} \frac{\|\alpha s\|^2}{\|\hat{s} - \alpha s\|^2} \quad (4)$$

here, $\alpha = \langle \hat{s}, s \rangle / \|s\|^2$ and $\alpha s$ represents the projection of the estimate onto the reference. Training minimizes the negative SI-SDR averaged over the two sources.

$$\mathcal{L} = -\frac{1}{2}[SI-SDR(\hat{s}_1, s_1) + SI-SDR(\hat{s}_2, s_2)] \quad (5)$$

## 3. EXPERIMENTAL SETUP

### 3.1. Dataset Generation

We generate spatialized three-second binaural mixtures (16 kHz) from LibriMix [12] speech and DNS noise [13], rendered with the RAZR room simulation toolkit [14] using B&K HRTFs [15]. Each scene contains two talkers at fixed azimuths ±60° between 1m and 2m, and one noise source at a random azimuth between 2m and 5m. We generate anechoic and reverberant mixtures with $T_{60}$ uniformly drawn from 0.1s to 0.6s; the listener is placed slightly off-center in a 12 m × 12.5 m × 3 m room. Noise is scaled to –10 dB to 20 dB SNR in 5 dB steps, and mixtures are peak-normalized to 0.99. Direct-path ear renders provide PIT-free targets (+60°→right ear, −60°→left ear). Splits use fixed seeds and a unique speaker pairs constraint to create 100 hours of training data and 10 hours of test and validation dataset.

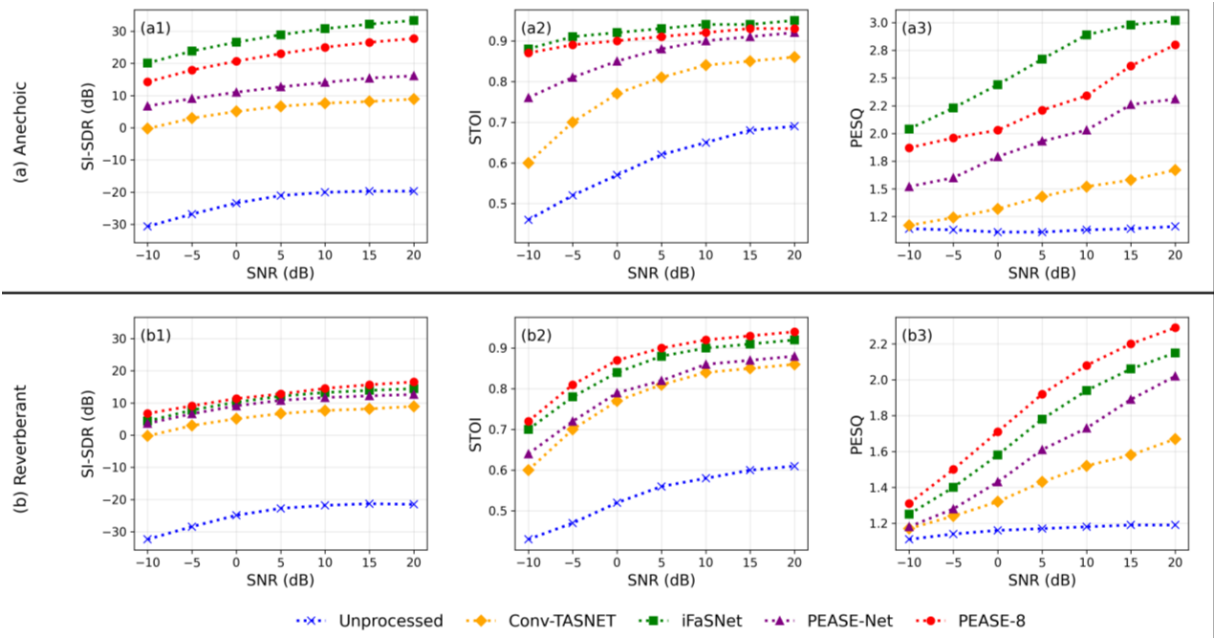

Fig. 2. SNR-stratified two-speaker results: (a) anechoic, (b) reverberant; SI-SDR (a1, b1), STOI (a2, b2), PESQ (c1, c2).

## 3.2. Training Setup

PEASE-8 is trained end-to-end on direct-path ear targets using negative SI-SDR loss. Complex STFTs (32 ms Hann, 50% overlap) are computed on-the-fly with identical iSTFT parameters for reconstruction. PIT is redundant since decoder heads are deterministically tied to its fixed-azimuth target via the ear-conditioned skip connections. Optimization uses Adam (learning rate = 1e-4), halving the learning rate if validation SI-SDR does not improve for five consecutive epochs. Training runs for up to 100 epochs, with early stopping triggered if validation SI-SDR fails to improve for ten consecutive epochs. A dropout of 0.1 is applied to the decoder layers for regularization.

## 3.3. Evaluation Metrics

We report the Scale-Invariant Signal-to-Distortion-Ratio (SI-SDR) [16], the Short-Time Objective Intelligibility (STOI) [17] score and the Perceptual Evaluation of Speech Quality (PESQ) [18]. All metrics are computed per source and averaged across the test set.

## 4. RESULTS AND DISCUSSION

We compare PEASE-8 against PEASE-Net [9] which provides a direct baseline, Implicit Filter-and-sum Network (iFaSNet) [5] (adapted to eight-channel inputs), and Multichannel Conv-TasNet [19] herein referred to as Conv-Tasnet. All baselines are in a non-causal setting and are retrained under identical conditions for fair comparison.

Table 1. Consolidated Separation performance in Anechoic and Reverberant conditions.

| Model | Anechoic | | | Reverberant ($T_{60}$) | | |
|---|---|---|---|---|---|---|
| | SISDR | STOI | PESQ | SISDR | STOI | PESQ |
| Unprocessed | -22.84 | 0.60 | 1.13 | -24.74 | 0.53 | 1.12 |
| Conv-Tasnet | 7.52 | 0.78 | 1.42 | 5.61 | 0.68 | 1.27 |
| iFaSNet | **27.9** | **0.92** | **2.61** | 10.85 | 0.84 | 1.74 |
| PEASE-Net | 12.1 | 0.86 | 1.92 | 7.56 | 0.79 | 1.59 |
| PEASE-8 | 22.18 | 0.91 | 2.26 | **12.37** | **0.87** | **1.86** |

### 4.1. Overall Performance Analysis

Table 1 summarizes absolute SI-SDR, STOI, and PESQ under anechoic and reverberant conditions. In anechoic conditions, iFaSNet attains the highest SI-SDR at 27.90 dB, exceeding PEASE-8 by 5.72 dB. Even so, PEASE-8 substantially outperforms PEASE-Net (12.10 dB, +10.08 dB) and the Conv-TasNet baseline (7.52 dB, +14.66 dB). In terms of intelligibility and perceptual quality, PEASE-8 records a STOI of 0.91 and a PESQ of 2.26.

Under reverberant conditions, PEASE-8 demonstrates clear superiority by achieving 12.37 dB SI-SDR, which is 1.52 dB higher than the next best model, iFaSNet (10.85 dB), and 4.81 dB above PEASE-Net (7.56 dB) and 6.7dB above Conv-TasNet (5.61dB). PEASE-8 also leads in intelligibility and perceptual quality, with a STOI of 0.87 (0.03 improvement over iFaSNet) and a PESQ of 1.86 (0.12 above iFASNet). These contrasting results within anechoic and reverberant conditions highlight PEASE-8's robustness to realistic environments.

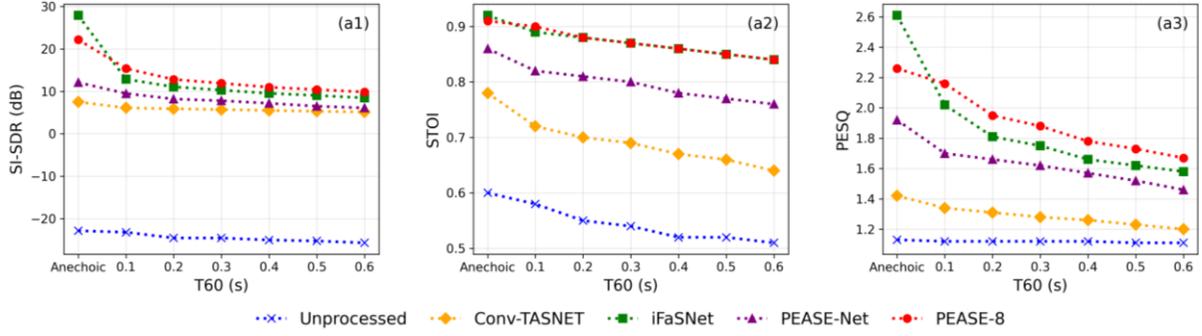

Fig. 3. T60-stratified two-speaker results: (a) anechoic, (b) reverberant; SI-SDR (a1, b1), STOI (a2, b2), PESQ (c1, c2).

### 4.2. Stratified Analysis

*4.2.1. SNR Analysis*

Figure 2 shows SI-SDR, STOI, and PESQ trends across SNR for both acoustic regimes. In the anechoic case, iFaSNet consistently leads, maintaining a nearly constant 5.6–5.9 dB SI-SDR margin over PEASE-8 from −10 to 20 dB SNR; both models improve by ~13 dB over this range, and STOI/PESQ rise in parallel. Under reverberation, the trend reverses: PEASE-8 exceeds iFaSNet by +2.40 dB at −10 dB SNR and remains ahead by ~0.7–2.1 dB across the remaining bins (peak margin at 20 dB SNR: +2.10 dB). It also delivers higher STOI and PESQ throughout.

*4.2.2. Reverberation Analysis*

Figure 3 shows SI-SDR, STOI, and PESQ across reverberation times—starting from the anechoic point through $T_{60}$= 0.6 s. In the anechoic case, PEASE-8 begins at 22.18 dB SI-SDR, 0.91 STOI, and 2.26 PESQ, outperforming PEASE-Net (12.10 dB, 0.86, 1.92) and Conv-TasNet (7.52 dB, 0.78, 1.42) but trailing iFaSNet (27.90 dB, 0.92, 2.61). As $T_{60}$ increases, all models degrade, yet PEASE-8 exhibits the slowest decline—from 22.18 dB to 9.85 dB SI-SDR, 0.91 to 0.84 STOI, and 2.26 to 1.67 PESQ—maintaining a 3 to 5 dB SI-SDR margin over baselines at each $T_{60}$. In contrast, iFaSNet falls from 27.90 dB to 8.46 dB, PEASE-Net from 12.10 dB to 6.10 dB, and Conv-TasNet from 7.52 dB to 5.12 dB, underscoring PEASE-8's superior ability to preserve separation fidelity and perceptual quality under increasing reverberation.

### 4.3. Discussion

The results highlight a clear contrast between performance in idealized versus realistic acoustic scenarios. In anechoic environments, iFaSNet excels, achieving 27.90 dB SI-SDR, 0.92 STOI, and 2.61 PESQ, while PEASE-8 closely follows with 22.18 dB, 0.91, and 2.26, significantly outperforming PEASE-Net (12.10 dB, 0.86 and 1.92) and Conv-TasNet (7.52 dB, 0.78 and 1.42). However, as reverberation increases, PEASE-8's ear-conditioned attention skips and fixed-azimuth decoders enable it to maintain robust separation fidelity, delivering 12.37 dB SI-SDR, 0.87 STOI, and 1.86 PESQ that is 2 to 4 dB better than the next best model, iFaSNet, and far above other baselines.

PEASE-8 is currently trained and evaluated only at fixed azimuth in a single simulated room, so generalization to other angles, moving speakers, and diverse rooms remains untested. Moreover, the model is noncausal, so real-time use requires a low latency/causal design. By broadening training to varied acoustic scenes and implementing an efficient causal multi-mic pipeline, PEASE-8 is well-positioned for practical deployment on ear-mounted platforms, delivering robust separation in everyday reverberant environments.

## 5. CONCLUSION

We presented PEASE-8, an eight-microphone, phase-aware separator with ear-conditioned outputs and a unique raw-STFT skip that bypasses the entire encoder pathway, enabling PIT-free training. PEASE-8 delivers superior separation and intelligibility in reverberation ($T_{60} \approx 0.6$s: 12.37 dB SI-SDR, 0.87 STOI, 1.86 PESQ) while remaining competitive anechoically (22.18 dB SI-SDR, 0.91 STOI, 2.26 PESQ). PEASE-8 delivers the best performance in reverberant conditions, while remaining competitive in anechoic settings. Stratified analyses show a shallower degradation with reverberation and consistent gains across SNR. Future work will address variable direction of arrival and low-latency deployment on resource-constrained devices.

## 6. ACKNOWLEDGEMENTS

This work is a part of Collaborative Research Centre SFB 1330 Hearing Acoustics (HAPPAA) - Project B2. This project is funded by the Deutsche Forschungsgemeinschaft (DFG, German Research Foundation) – Project-ID 352015383 – SFB 1330. Simulations were conducted on the HPC cluster ROSA funded by the DFG under INST 184/225-1 FUGG.

## 7. COMPLIANCE WITH ETHICAL STANDARDS

This is a computational study using publicly available corpora (LibriSpeech and DNS Noise) for which no ethical approval was required.